\documentclass[12pt,a4paper,doublespace]{article}
\usepackage[english]{babel}
\usepackage[dvips]{graphics}
\usepackage{epsfig}
\usepackage[latin1]{inputenc}
\usepackage{amssymb}
\usepackage{indentfirst}

\begin{document}

\begin{center}
{\bfseries Seismic Activity in Tucunduba Reservoir, Senador S\'a-CE, Brazil, 1997-1998.} 

George Sand Fran\c{c}a\footnote{george@dfte.ufrn.br}, Joaquim Mendes Ferreira \\ and Mario Koechi Takeya \\
Departamento de F\'{\i}sica Te\'orica e Experimental, \\
Universidade Federal do Rio Grande do Norte, \\
Natal-RN, 59072-970, Brazil

\end{center}

\begin{abstract}

The Tucunduba Dam is about 290 km West of Fortaleza, Cear\'a state. The seismic monitoring of the area, with a local network, has began on June 11, 1997, soon after the occurrence of the event with magnitude 3.2 $m_b$ on June 09, 1997. The monitoring was done with one the analogical station (used for magnitude duration measurement and statistical control activity) and seven digital stations. The digital stations, with three components each, were operated from June to November 1997. In this work, the data collected during digital monitoring was analyzed to determine hypocenters and focal mechanism. To determine hypocenters, the HYPO71 program was used, with half-space model, whose parameters were 5.95 km/s, for P-wave velocities, and 1.69 for the ratio between P and S-wave velocity. The active zone was nearly 1 km length, with depth between 4.5 and 5.2 km. With 16 events recorded in the same six stations, we determined the direction of the fault plane (NE-SW). The fault mechanism is strike-slip with a small normal component. The dip estimate was 65$^{\circ}$SE with FPFIT and 80$^{\circ}$SE with FOCMEC. Preliminary estimates of maximum horizontal compressive stress, using P-axis direction, were in accordance with Ferreira {\it et al.}(1998). The small difference is probably due to influence of the sedimentary basin in the regional stress. The active area is in accordance with seismicity described by Assump\c{c}\~ao (1998), i. e., the combination regional stress, local flexural effect from thick sedimentary loads, and a presumably weaker crust from Mesozoic thinning.
\end{abstract}

{\it Key words: Seismicity, focal mechanism}

\date{}

\newpage

\section{Introduction}

	The Northwest of Cear{\'a} is an area of seismic activity known since the XIX century (Ferreira \& Assump\c{c}\~{a}o, 1983; Berrocal {\it et al.}, 1984; Ferreira {\it et al.}, 1998). On June 09, 1997 (district of Serrota, Senador S\'a-CE), an event with magnitude $m_b = 3.2$ (Intensity III-IV MM) happened, scaring the local population. The growth of the seismic activity in Serrota, specially in the Tucunduba Reservoir, led the UFRN to study the seismic activity, to determine the hypocenter and displacement mechanism with large precision and to try to understand the local seismicity.

	For this study, an analogical station and a digital network with seven stations were installed (Figure \ref{fig:stations}). The analogical station operated from June 11, 1997 to August 14, 1998 and the digital network from June 18 to November 05, 1997. The analogical station was used in the statistics of the earthquakes as well as to obtain the magnitude with the duration, using the same parameters presented by Blum \& Assump\c{c}\~{a}o (1990), since the analogical station operated in the same conditions.
\vspace{-0.2 cm}

\section{Geological comments}

	The seismic area is part of the Borborema province which is divided in five tectonic domains (Brito Neves {\it et al.}, 2000). The Northwest of Cear{\'a} is divided by two of these domains; the Medium Corea\'{u} Domain (MCO) and the Cear\'{a} Central Domain (CC).

\begin{figure}[p]
\begin{center}
{\scalebox{0.7}{\includegraphics{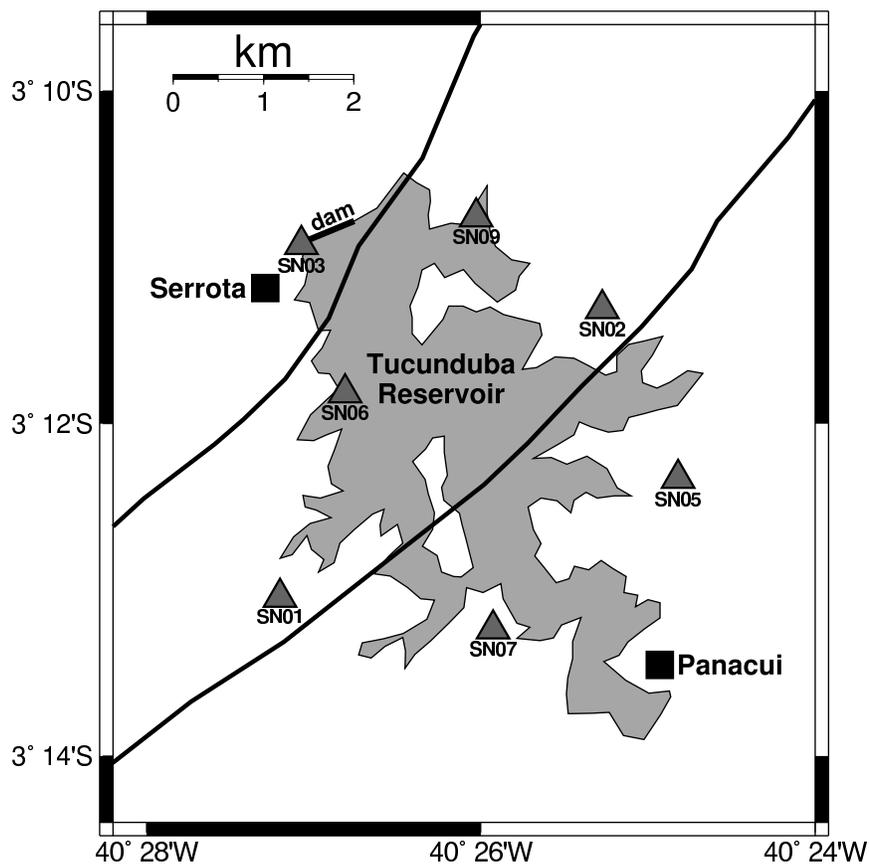}}}
\end{center}
\caption{\small Map of the locations of the seismographic stations. The triangles represent the seismographic stations. In the site SN01 were deployed an analogical (SN1A) and a digital station.
The squares represent the small villages; lines crossing the dam are faults (DNPM, 1973).}
\label{fig:stations}
\end{figure}

\begin{figure}[p]
\begin{center}
{\scalebox{0.7}{\includegraphics{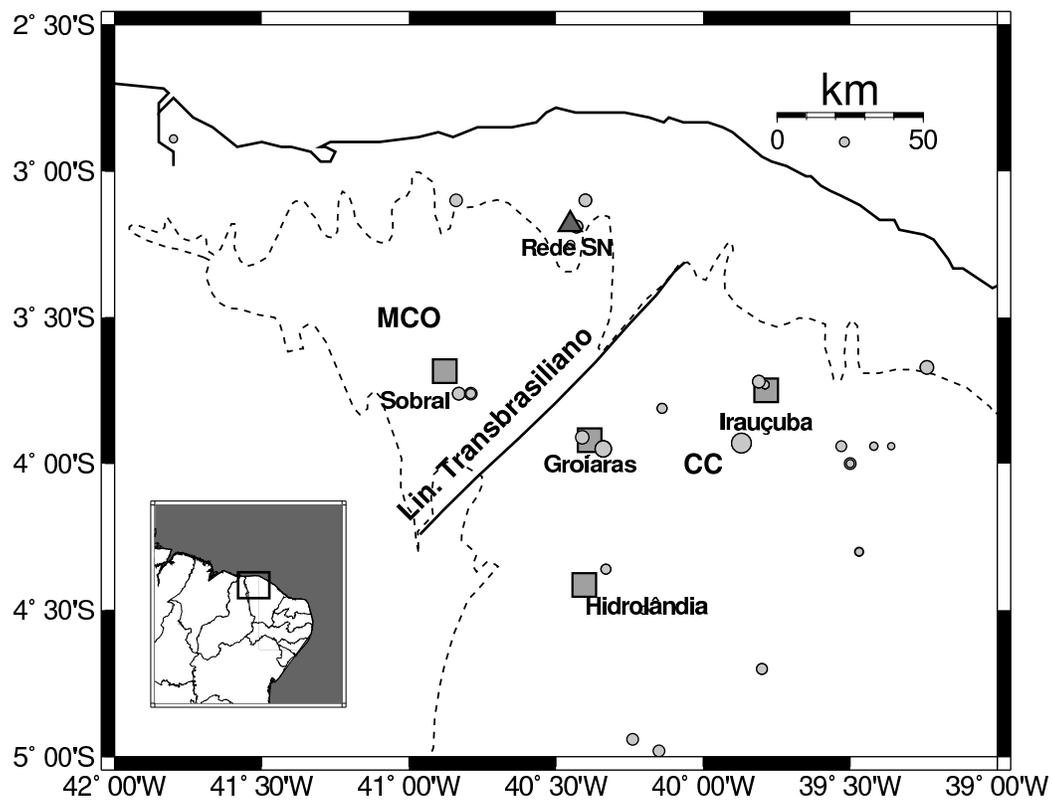}}}
\end{center}
\caption{\small Map separating the two main domains NW of Cear\'{a}. The Medium Corea\'{u} Domain = MCO, the Cear\'a Central Domain = CC  and transbrasiliano lineament. Cicles are epicenters recorded by UFRN network.}
\label{fig:geologia}
\end{figure}

\begin{figure}[p]
\begin{center}
\rotatebox{-90}{\scalebox{0.9}{\includegraphics{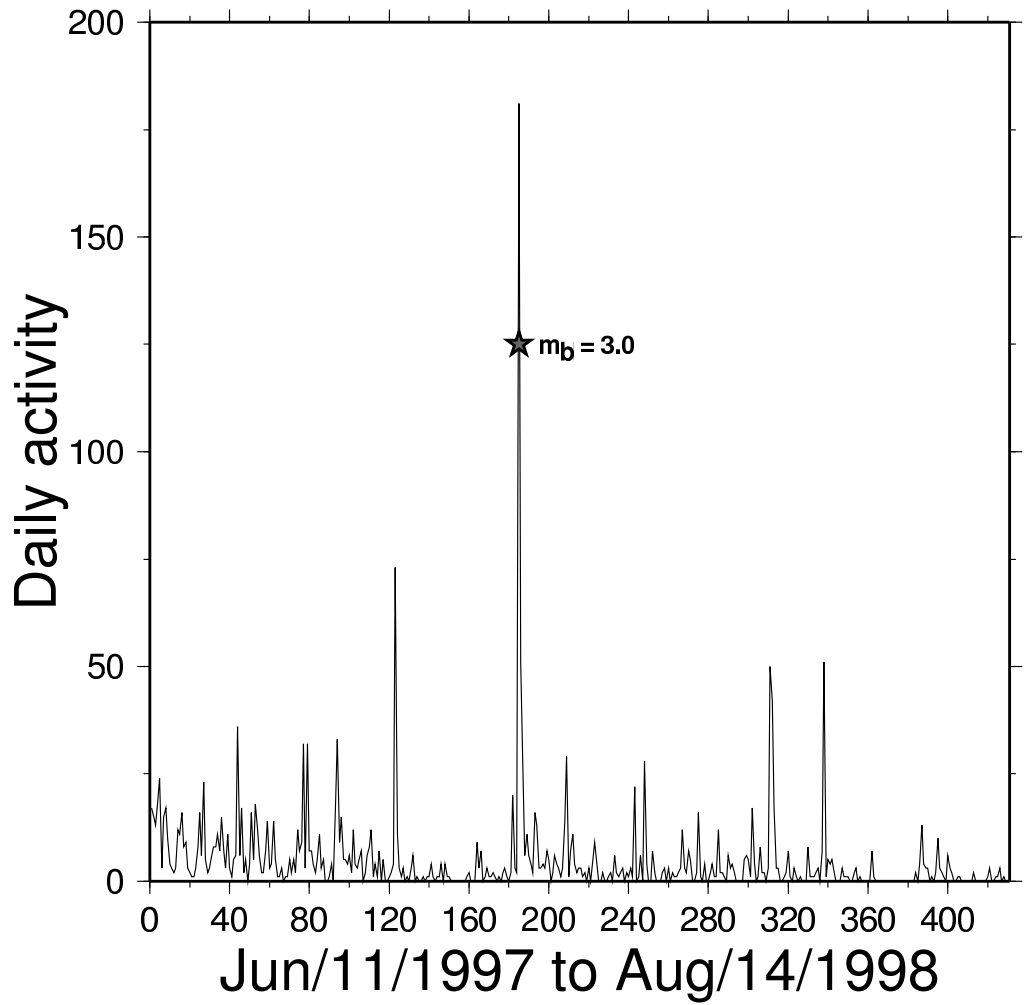}}}%histogram.eps
\end{center}
\caption{\small Tucunduba activity, Number of daily events recorded by SN1A station from June 1997 to August 1998.}
\label{fig:histograma}
\end{figure}

\begin{figure}[p]
\begin{center}
\rotatebox{-90}{\scalebox{0.65}{\includegraphics{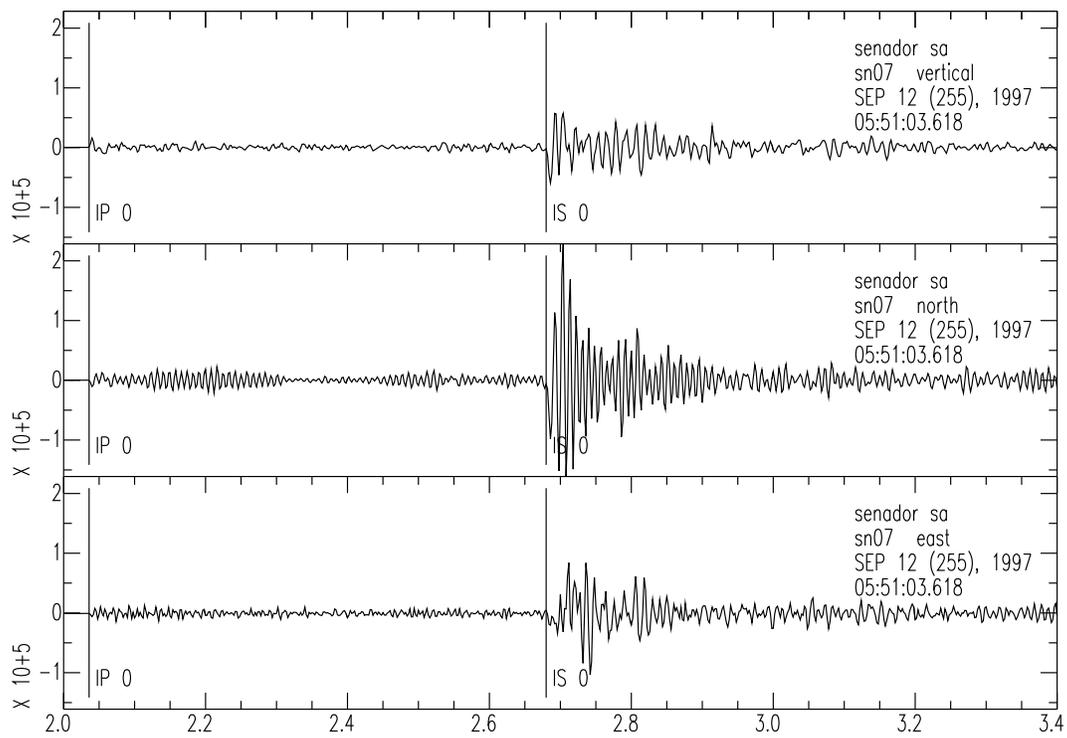}}}
\end{center}
\caption{\small Seismogram recorded at SN07 station with clear P and S arrivals in the vertical component. Event n$^{\circ}$ 15 (table \ref{tab:hypo}).}
\label{fig:sismo}
\end{figure}

\begin{figure}[p]
\begin{center}
{\scalebox{0.7}{\includegraphics{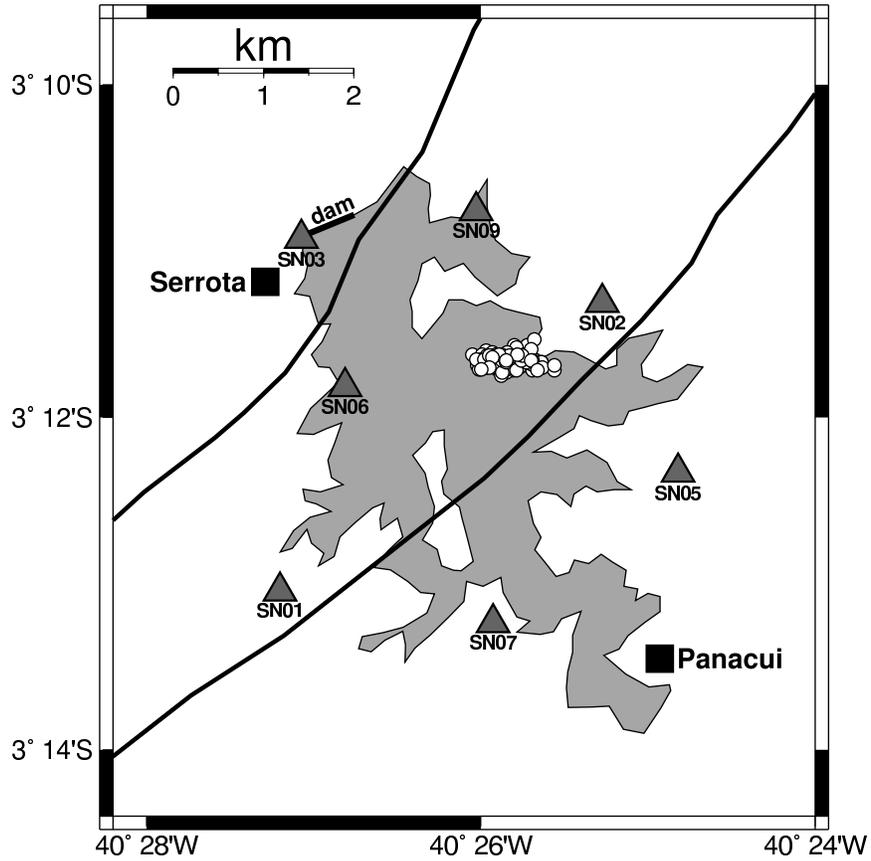}}}
\end{center}
\caption{\small Map of Tucunduba dam epicenters, with 160 selected events (circles), recorded by at least four stations (rms $\leq$ 0.02s, erh $\leq$ 0.02 km, erz $\leq$ 0.2 km). Triangles indicate seismic stations.}
\label{fig:epitotal}
\end{figure}

\begin{figure}[p]
\begin{center}
\scalebox{0.7}{\includegraphics{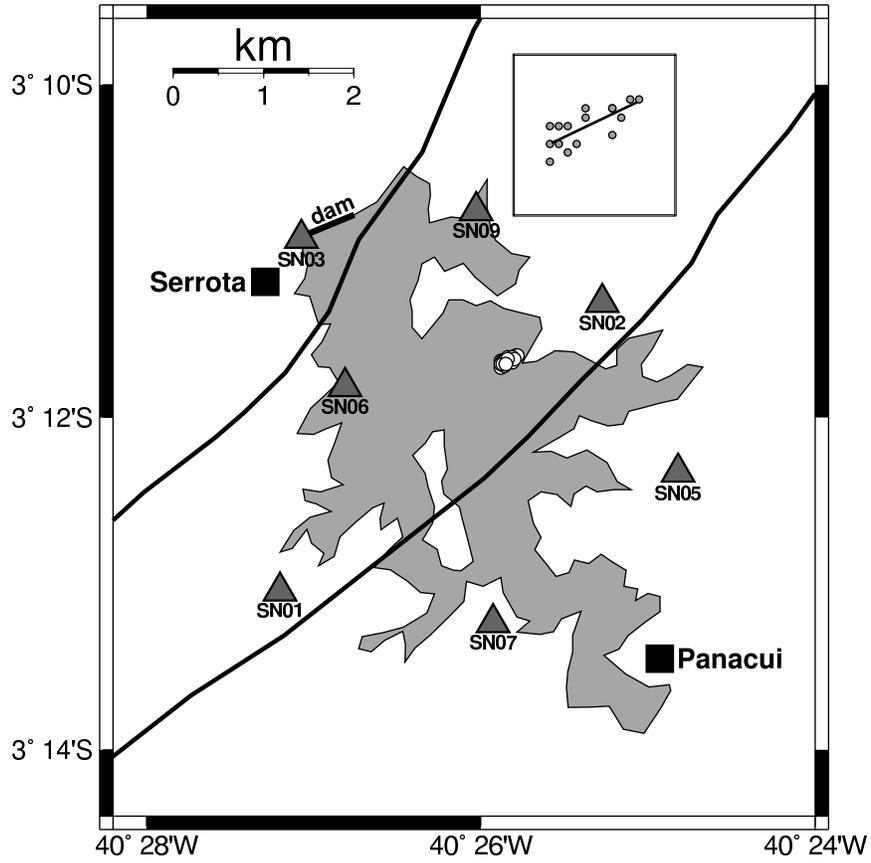}}
\end{center}
\caption{\small Map of epicenters with 16 selected events (circles), recorded by at least six stations. Triangles indicate seismic station. Small squares represented a closer look of the picenters.}
\label{fig:epi12}
\end{figure}

\begin{figure}[p]
\begin{center}
\scalebox{0.7}{\includegraphics{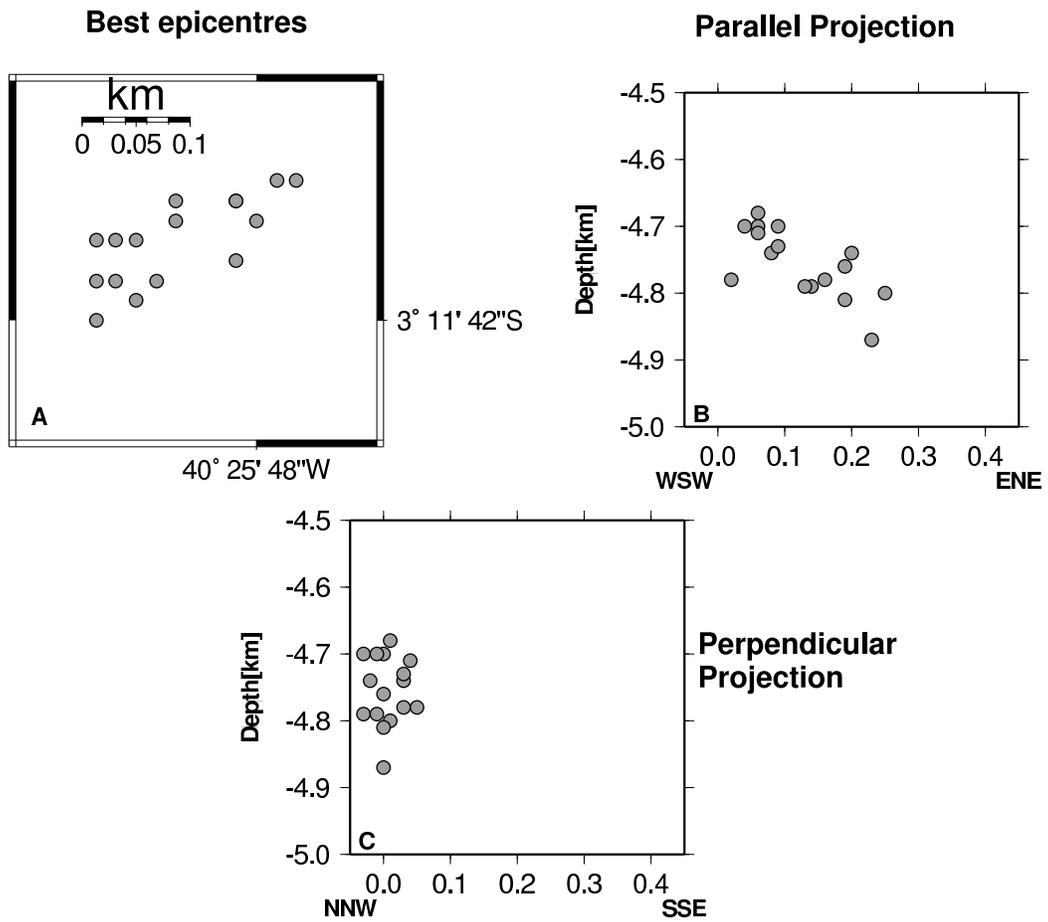}}
\end{center}
\caption{\small A. Detailed map of epicenters in Figure 6. B e C. Projections of the hypocenters on the vertical plane across the probable fault plane (Parallel and Perpendicular), respectively. Events used to estimate fault plane.}
\label{fig:detail}
\end{figure}

	The MCO Domain is situated between the margin of the S\~ao Luis-West Africa craton and the Transbrasiliano/Kandi Lineament. The domain consists of basement (2.5 Ga) high-grade methamorphic rocks and pelitic-carbo-nate fold belts. Along the Transbrasiliano lineament occured a series of transtensional basin transpressional and post-orogenic plutons occupying pull-apart; many of these are covered by the Phanerozoic sedimentary rocks of the Parna{\'{\i}ba Basin (Gusm\~ao, 1998).

	The CC domain is situated between the Transbrasiliano/Kandi lineament and the Senador Pompeu lineament (Figure \ref{fig:geologia}). This domain consists of gneissic basement formed during the Transamazonian collage, with invulsion of an Archean nucleus. The basement has 2.2 and 2.1 Ga. The domain also contains a series of middle Neoproterozoic supracrustal sequences or remanent of folds belts and expressive Brasiliano plutonism.

	The dam is located in the bound of the MCO domain and coverings Phanerozoic. Close to the dam, there is a sucession of horst and grabens (Torquato \& Nogueira Neto, 1996).

\section{Seismic Activity}

	The large event recorded in the Tucunduba Dam, with magnitude 3.2 $m_b$, was on June 9th, 1997 and was felt mainly in Serrota, a small villages of Senador S\'a, by the margins of the dam (Figure \ref{fig:stations}), where roofs were shaken, glasses of stores balanced, school materials were dropped off desks and, the inhabitants were afraid (Fran\c{c}a, 1999). Then, on June 11th, 1997, the analogical seismograph station, with MEQ-800 smooked paper recorder (SN1A), was installed. Seven digital short period portable stations were also installed, on June 18, 1997. Except the station SN03, all the stations were installed in granitic/gneissic bedrocks of the crystalline basement, allowing low noise and clear P and S wave arrivals. The digital stations clocks were corrected once each hour with GPS and P-wave readings had accuracy to $\pm 0.001 s$.

	The station SN1A operated from June 11, 1997 to August 14, 1998, recording a total of 2217 events (Figure \ref{fig:histograma}), and the largest magnitude was $m_b$ = 3.0 on December 12, 1997. 

\subsection{Determination of Hypocenters}

	Hypocenters were determined with the program HYPO71 (Lee \& Lahr, 1975). The half-space velocity model has been producing good results when applied to several areas of Brazilian Northeast (Ferreira {\it et al.}, 1987; 1995; 1998; Takeya {\it et al.,} 1989; do Nascimento, 1997). This half-space model was adopted, since the seismic area is in the precambrian basement composed of consolidated rocks and low attenuation, generating clear P and S waves arrivals in the seismographs (Figure \ref{fig:sismo}).

	Various tests were made to find the best P velocity ($v_P$) and ($v_P/v_S$) ratio. We varied  P wave velocity between 5.4 and 6.45 km/s and $v_P/v_S$ ratio between 1.60 and 1.74. The velocity model with the lowest root mean square (rms) time residual had $v_P/v_S$ = 1.69 and $v_P$ = 5.95 km/s. The value for $v_P$ is acceptable, since in different parts of the world, the $v_P$ in granitic/gneissic upper crustal rocks varies from 5.8 to 6.4 km/s (Christensen, 1982; Christensen \& Mooney, 1995). For this model, all 160 selected events (recorded by at the least four stations) had rms residual 0.03 sec and horizontal errors less than 0.5 km. The most of the events have rms $\leq$ 0.01, erh $\leq$ 0.1 km and erz $\leq$ 0.1 km, constitutes a set of weel determinated.
This is basically due the site of deployment (mainly bedrock), the stations near the epicentral area anda sample rate of 500 sample per second.

	Figure \ref{fig:epitotal} shows the epicenters of the events with rms residuais less than 0.03 sec and horizontal errors less than 0.5 km, recorded during the digital network operation. As can be seen, all events occurred inside the lake, with about 1 km length and depths between 4.5 and 5.2 km. We do not know the effect of the water in the trigger, since the dam is very old, shallow and seismic activity is not known previous

\section{Fault plane}

	To estimate the fault plane, we used 16 well-located events recorded at the same six stations (Table \ref{tab:hypo}, Figure \ref{fig:epi12}). The hypocenters were distributed in 0.3 km length, with depth varying from 4.68 to 4.87 km (Figure \ref{fig:detail}).

	Starting from these data, we determined, by least squares, the direction and the dip of the fault plane. The azimuth was 60$^{\circ} \pm 10^{\circ}$ and the dip 88$^{\circ}\pm 11^{\circ}$. Projections of the hypocenters used in the calculation, in parallel and perpendicular plans to the fault plan, are shown in the Figure \ref{fig:detail}. The vertical and horizontal maximum errors observed for that set of data were 0.1 km, while the range of the active zone was of 0.3 km and the variation in depth was of 0.2 km. In these conditions, the hypocentral data can  provide only a rough estimate of the fault plane orientation.

\section{Focal mechanisms}

	Composite focal mechanisms were determined using clear P wave first motions with aid of the FPFIT program (Reasenberg \& Oppenheimer, 1985). FPFIT program does a grid search to find the solution that minimizes a weighted sum of discrepancies in the polarities, considering both the estimated variance of the data and the theoretical P wave radiation amplitude (Reasenberg \& Oppenheimer, 1985). The individual focal mechanisms were determined using S/P amplitudes ratios in additition to P wave first motions, according to the method of Kisslinger (1980) and Kisslinger {\it et al.} (1982). For individual focal mechanism, we used the FOCMEC program (Snoke {\it et al.}, 1984), which does a grid search of the nodal parameters to find the solution the best fits the observed log (amplitude ratio).

	Firstly, with the 16 events of Table \ref{tab:hypo} we used the fault plane obtained by least square with reference, we estimated the composite focal mechanism (with number of $P+S$ readings $\geq$ 12 and at least five clear P onset). Added the short-period station located in Sobral-CE (SB17), epicentral distance of 85.5 km, for the two largest events in Table \ref{tab:hypo} (Event 1 and 2). To determine the focal mechanism by using FPFIT, we limited the fault plane near 60$^{\circ}$ with range 30$^{\circ}$ and result of the composite focal mechanism (azimuth, dip and the rake), shown in the Table \ref{tab:FPFIT}--Figure \ref{fig:fpfit}. The solution of the focal mechanism is approximately a NE-SW fault, with strike-slip dextral movement with normal component. The fault plane was chosen by taking into account the hypocentral distribution (Figure \ref{fig:epi12}). The composite focal mechanism solution is a strike-slip type, which is predominant in the northeast of Brazil (Assump\c{c}\~ao {\it et al.}, 1985; 1989; Ferreira {\it et al.,} 1995; 1998). The direction of the P axis (278$^{\circ}$), obtained by FPFIT, differs from 6$^{\circ}$ of the maximum horizontal compressive stress ($SH_{max}$), obtained by Ferreira {\it et al.} (1998) in NE Cear\'a region.

	Secondly, seven events of Table \ref{tab:hypo} were selected (Table \ref{tab:FOCMEC}), by using as selection criterion the largest number of clear polarities of the S wave. The results obtained through the FOCMEC are in Table \ref{tab:FOCMEC}, limiting the range of orientation and dip of the axes B (null axis), and A and N (correspond to Herrmann's X and Y axes), with step 2$^{\circ}$. The events are numbered in accordance with Table \ref{tab:hypo} and the best solutions of focal mechanisms with its P and T axes, for the reliable mechanisms for FOCMEC. The results showed a predominance of strike-slip mechanism (1, 11, 14, 15 and 16). Event 6 has normal mechanism and event 13 a normal with strike-slip component. Without considering the dip, several mechanisms have close direction to the estimated azimuth with previous methods (60$^{\circ}$)(Table \ref{tab:FOCMEC}).

	The average strike for all events was 65$^{\circ}$. The average dip was 80$^{\circ}$. These average values were close the values obtained by fitting of the fault plane by least squares and composite focal mechanisms (FPFIT).

	To estimate $S_{Hmax}$ with individual focal mechanism, we did an average of the P-axes directions (for the events of mechanism strike-slip) and B-axes (for the events of normal mechanism). The obtained value was 258$^{\circ}$, between the values obtained by FPFIT for direction of the P axis (278$^{\circ}$) and the value of $S_{Hmax}$ for the northwestest Cear\'a (293$^{\circ}$) obtained by Ferreira {\it et al.} (1998).

\section{Discussion} 

	A network composed by seven portable digital stations was used to monitoring the seismic activity in the Tucunduba Dam, almost all installed in granitic/gneissic bedrocks of the crystalline basement. It was possible to obtain records of good quality, with clear P and S waves. Besides, the network configuration has two stations that were installed close to the epicentral area. Thus, it was possible to determine the hypocenters with good accuracy.

	The active zone was too small, less than about 1 km length and depth varying from 4.5 to 5.2 km, considering all events analyzed. It was inside the Tucunduba Reservoir. 

	The set of 16 events recorded by the same six stations was selected from the method of the least square, obtaing values of 60$^{\circ}$ for the azimuth and 88$^{\circ}$ for the dip. 

	The focal mechanism solution shows a NE-SW fault, with strike-slip dextral movement with normal component. The fault plane was chosen by taking into account the hypocentral distribution (Figure 6). The direction of the axes P (278$^{\circ}$), obtained by composite focal mechanism, differs 15$^{\circ}$ of the direction of the horizontal maximum stress ($S_{Hmax}$), obtained by Ferreira {\it et al.} (1998). In the determination of individual focal mechanisms, there is three or four stations with the same polarities, which may be indicated that mechanisms is not change, thus the strike-slip mechanism (four) is fully with expected results. The average direction of the individual planes ranges from 65$^{\circ}$ to the azimuth and 80$^{\circ}$ for the dip, which are close to the values obtained by the composite mechanism and least squares. As the direction of horizontal maximum stress ($S_{Hmax}$) was 258$^{\circ}$, it differs 6$^{\circ}$ of the best estimated by Ferreira {\it et al.} (1998).

	The seismic activity is located in the border of the basin, common characteristic of the seismicity of the Northeast (Assump\c{c}\~ao, 1998; Ferreira {\it et al.}, 1998). The focal mechanism and fault plane estimated (the strike-slip type) is predominant in the Northeast of Brazil (Assump\c{c}\~ao {\it et al.}, 1985; 1989; Ferreira {\it et al.}, 1995; 1998). The $S_{Hmax}$ estimated has a small difference. Probably, that difference is because Ferreira {\it et al.}, (1998) used events to the South of the active area and that the line of the coast suffers more flexural effect than area to the South.

	The NW area of Cear\'a was analyzed up to now with events that happened on the basements, the seismic area of the Tucunduba reservoir is consistent with stress regime (Ferreira {\it et al.}, 1998}), in other words, $S_{Hmax}$ parallel the Northeast coast, where this active is of the border of the boundary basin.
	
\vspace{0.3 cm}

%{\bfseries Acknowledgments}

%We specially thank Eduardo Menezes for his effort and responsibilities during field work. This work was supported by Brazilian Grants CNPq, FUNPEC/CONNECIT and FINEP. I thank Marcelo Assump\c{c}\~ao for discussions in the interpretation and data analysis. Maps were plotted using GMT (Wessel \& Smith, 1991) and seismograms were analysed with SAC (Goldstein {\it et al.,} 1991).

\newpage

\listoftables

\newpage

\begin{table}
\caption{Events recorded by the same the six stations (SN01, SN02, SN04, SN05, SN06, SN07 and SN09). H-origin is the origin hour, m$_D$ is the duration magnitude using the same parameters presented by Blum \& Assump\c{c}\~ao (1990) ($m_D = 1.64 log D - 0.97$).}
\begin{center}
\small{
\centering
\begin{tabular}{rrlllllc} 
\hline
\\
\multicolumn{1}{c}{DATA}&
\multicolumn{2}{c}{H-origin}&
\multicolumn{1}{c}{lat S}&
\multicolumn{1}{c}{long W}&
\multicolumn{1}{c}{depth}&
\multicolumn{1}{c}{$m_D$}
\\
\hline
\\
1&970904&2052&47.87&3-11.63&40-25.79&4.87&1.5\\
2&970905&0113&50.93&3-11.64&40-25.81&4.81&1.0\\
3&970905&0114&02.56&3-11.63&40-25.78&4.80&1.0\\
4&970909&0427&55.67&3-11.70&40-25.88&4.78&1.5\\
5&970911&2146&59.60&3-11.66&40-25.88&4.70&$<1.0$\\
6&970912&0517&31.43&3-11.66&40-25.87&4.74&$<1.0$\\
7&970912&0517&59.47&3-11.64&40-25.84&4.79&$<1.0$\\
8&970912&0518&09.73&3-11.66&40-25.86&4.70&$<1.0$\\
9&970912&0519&30.63&3-11.68&40-25.88&4.70&$<1.0$\\
10&970912&0519&40.12&3-11.67&40-25.81&4.78&$<1.0$\\
11&970912&0519&55.12&3-11.69&40-25.86&4.71&$<1.0$\\
12&970912&0523&55.15&3-11.68&40-25.87&4.68&2.2\\
13&970912&0536&47.90&3-11.64&40-25.81&4.76&1.5\\
14&970912&0548&26.82&3-11.65&40-25.80&4.74&1.5\\
15&970912&0551&04.71&3-11.65&40-25.84&4.79&2.3\\
16&970912&0843&28.86&3-11.68&40-25.85&4.73&$<1.0$\\
\hline
\end{tabular}}
\label{tab:hypo}
\end{center}
%\end{table}
%\begin{table}[p]
\caption{Result of the composite focal mechanism using FPFIT program (Reasemberg \& Oppenheimer, 1985).}
\begin{center}
\begin{tabular}{rllllcc} 
\hline
\\
\multicolumn{1}{c}{Azimuth}&
\multicolumn{1}{c}{dip}&
\multicolumn{1}{c}{Rake}&
\multicolumn{2}{c}{P-Az/plunge}&
\multicolumn{2}{c}{T-Az/plunge}\\
\hline
\\
$60^{\circ}\pm 11^{\circ}$&$65^{\circ}\pm 5^{\circ}$&$-174^{\circ} \pm 
4^{\circ}$&$278^{\circ}$&$27^{\circ}$&$15^{\circ}$&$14^{\circ}$\\
\hline
\end{tabular}
\end{center}
\label{tab:FPFIT}
\end{table}

\begin{table}[t]
\caption{Results of focal mechanism - FOCMEC. Each solutions with its mechanism focal and teh events correspond Table \ref{tab:hypo}.}
\begin{center}
\begin{tabular}{cccrccc} 
\hline
\\
\multicolumn{1}{c}{Events}&
\multicolumn{1}{c}{Azimuth}&
\multicolumn{1}{c}{dip}&
\multicolumn{1}{c}{Rake}&
\multicolumn{1}{c}{Mechanism}& 
Az./plunge & Az./plunge 
\\
 & & & & & eixo-P & eixo-T \\
\hline
1&44$^{\circ}$&68$^{\circ}$&-166$^{\circ}$&strike-slip&265$^{\circ}$/25$^{\circ}$&358$^{\circ}$/06$^{\circ}$\\
6&74$^{\circ}$&44$^{\circ}$&-84$^{\circ}$&normal&84$^{\circ}$/86$^{\circ}$&340$^{\circ}$/01$^{\circ}$\\
11&78$^{\circ}$&85$^{\circ}$&174$^{\circ}$&strike-slip&125$^{\circ}$/00$^{\circ}$&35$^{\circ}$/08$^{\circ}$\\
13&76$^{\circ}$&65$^{\circ}$&-45$^{\circ}$&normal+strike-slip&215$^{\circ}$/49$^{\circ}$&315$^{\circ}$/8$^{\circ}$\\
14&89$^{\circ}$&84$^{\circ}$&-10$^{\circ}$&strike-slip&45$^{\circ}$/12$^{\circ}$&135$^{\circ}$/03$^{\circ}$\\
15&51$^{\circ}$&72$^{\circ}$&-172$^{\circ}$&strike-slip&263$^{\circ}$/33$^{\circ}$&360$^{\circ}$/11$^{\circ}$\\
16&44$^{\circ}$&73$^{\circ}$&170$^{\circ}$&strike-slip&89$^{\circ}$/07$^{\circ}$&358$^{\circ}$/07$^{\circ}$\\
\hline
\end{tabular}
\end{center}
\label{tab:FOCMEC}
\end{table}

%\begin{table}[t]
%\\
%\caption{Results of axes P and T - FOCMEC (Snoke \it et al \rm,1984).}
%\begin{tabular}{rcrrrr} 
%\hline
%\\
%\multicolumn{2}{c}{Events}&
%\multicolumn{1}{c}{P-Strike}&
%\multicolumn{1}{c}{P-Plunge}&
%\multicolumn{1}{c}{T-Strike}&
%\multicolumn{1}{c}{T-Plunge}\\
% &mmdd-hhmm&&&&\\
%\hline
%1&09/04-2052&265$^{\circ}$&25$^{\circ}$&358$^{\circ}$&6$^{\circ}$\\
%6&09/12-0517&84$^{\circ}$&86$^{\circ}$&340$^{\circ}$&1$^{\circ}$\\
%?&09/12-0519&125$^{\circ}$&0$^{\circ}$&35$^{\circ}$&8$^{\circ}$\\
%13&09/12-0536&215$^{\circ}$&49$^{\circ}$&315$^{\circ}$&8$^{\circ}$\\
%14&09/12-0548&45$^{\circ}$&12$^{\circ}$&135$^{\circ}$&3$^{\circ}$\\
%15&09/12-0551&263$^{\circ}$&33$^{\circ}$&360$^{\circ}$&11$^{\circ}$\\
%3&09/12-0114&89$^{\circ}$&7$^{\circ}$&358$^{\circ}$&7$^{\circ}$\\
%\hline
%\end{tabular}\end{center}
%\label{tab:FOCMEC2}
%\end{table}

\newpage

\newpage

\listoffigures 

\newpage

\begin{figure}[p]
\begin{center}
{\scalebox{0.7}{\includegraphics{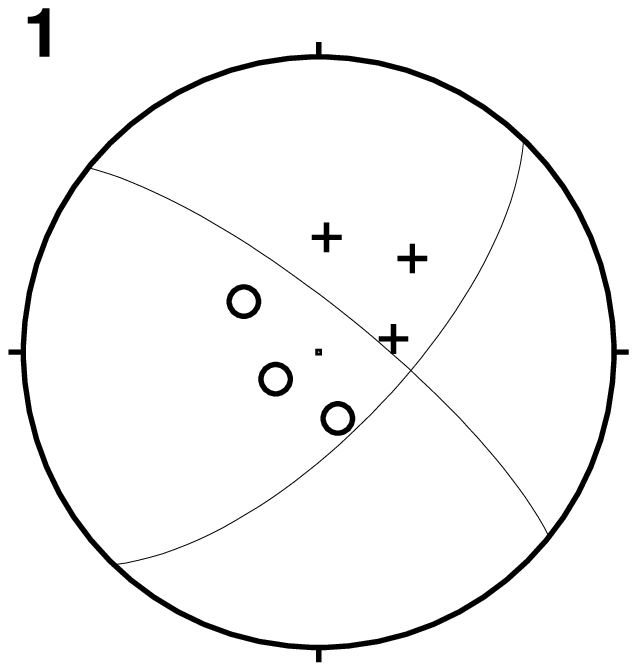}}}
{\scalebox{0.7}{\includegraphics{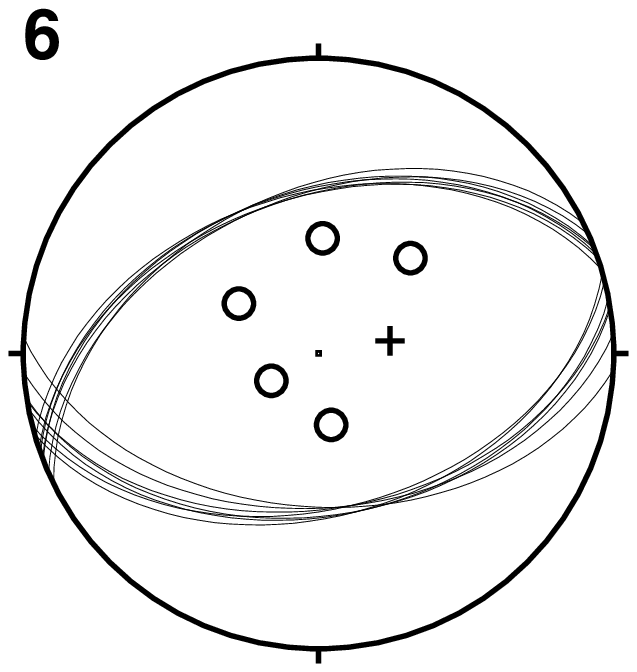}}}
{\scalebox{0.7}{\includegraphics{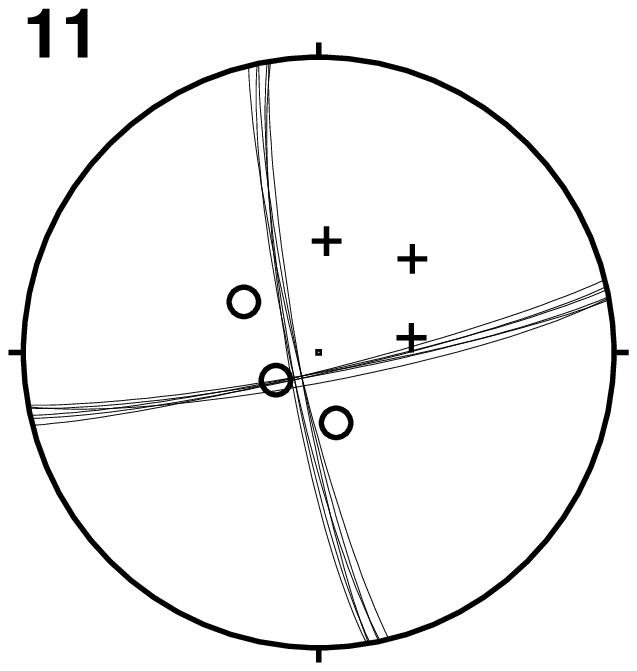}}}
{\scalebox{0.7}{\includegraphics{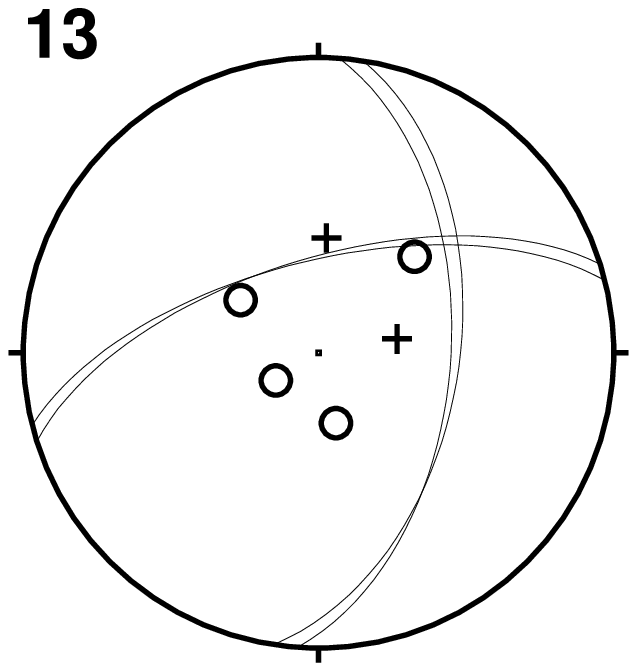}}}
{\scalebox{0.7}{\includegraphics{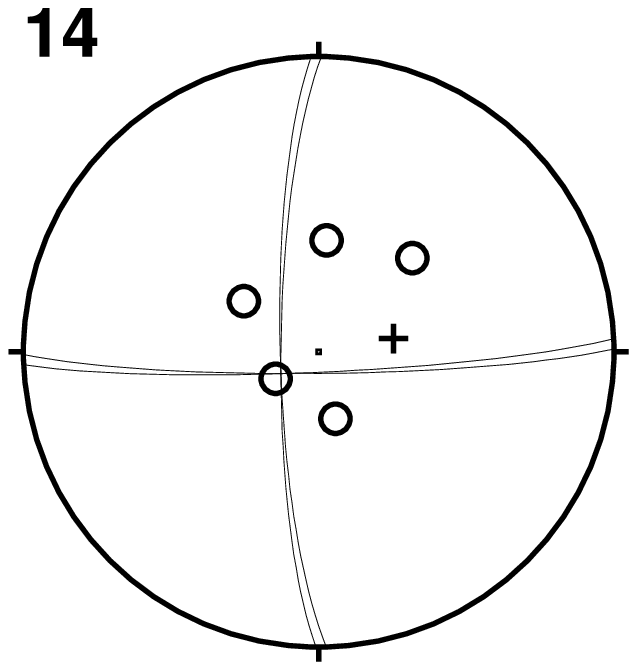}}}
{\scalebox{0.7}{\includegraphics{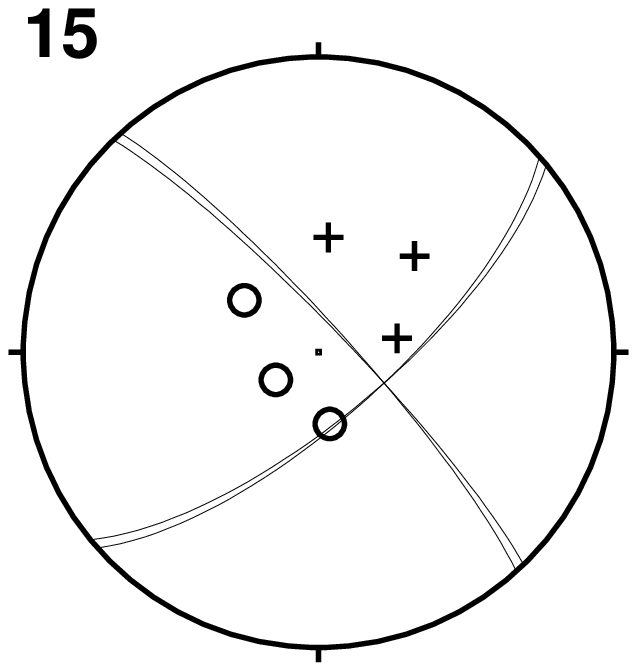}}}
{\scalebox{0.7}{\includegraphics{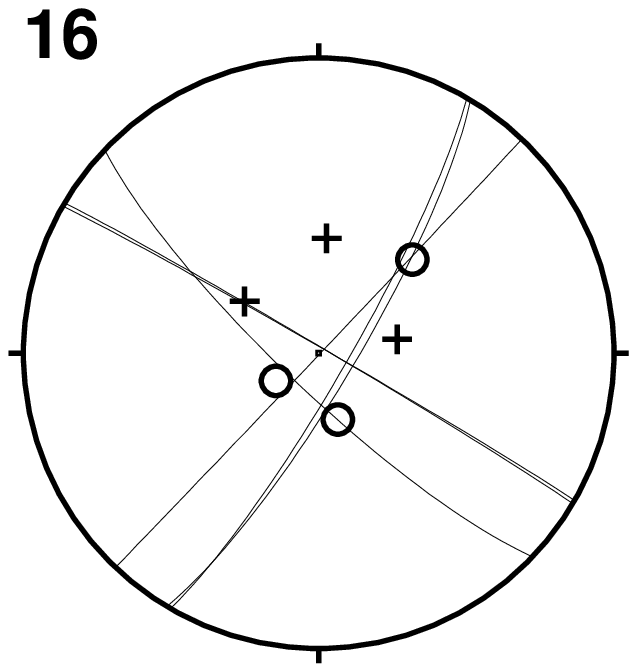}}}
\end{center}
\caption{\small Focal mechanisms using FOCMEC and half-space model; + = up and $\circ$ = down; Numbers refer to table \ref{tab:hypo}. }
\label{fig:focmec}
\end{figure}

\begin{figure}[p]
\begin{center}
{\scalebox{1}{\includegraphics{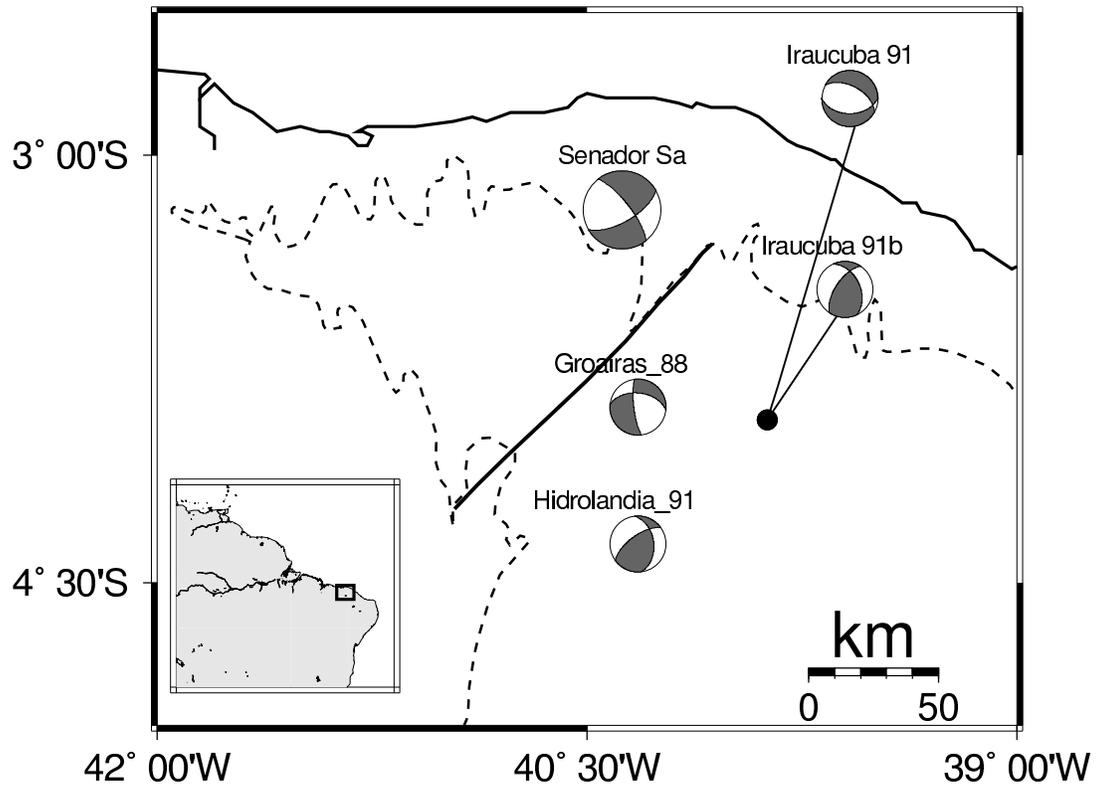}}}
\end{center}
\caption{\small Composite Focal Mechanism for 16 events (table 1). Lower hemisphere, equal-area projections. Black and Ash Circles denote compressive and extensive first motions dominated, respectively.}
\label{fig:fpfit}
\end{figure}

\end{document}